\documentclass{article}
\usepackage{spconf,amsmath,graphicx}
\usepackage{wrapfig}
\usepackage{tabularx}
\usepackage{booktabs}
\usepackage{textcomp}
\usepackage{microtype}
\usepackage[separate-uncertainty=true, multi-part-units=single]{siunitx}
\usepackage[belowskip=0pt,aboveskip=6pt]{caption}
\usepackage{bm}
\usepackage{subcaption}
\usepackage[hidelinks]{hyperref}
\usepackage{enumitem}
\usepackage{xcolor}
\usepackage{soul}
\usepackage{setspace}
\usepackage{wrapfig}
\usepackage{float}
\usepackage{filecontents}
\usepackage{etoolbox}
\usepackage{multirow}
\usepackage{glossaries}
\DeclareSIUnit\px{px}
\DeclareSIUnit\db{dB}
\usepackage[utf8]{inputenc}
\usepackage[style=ieee,backend=biber, url=false,doi=false,isbn=false,eprint=false, maxbibnames=200,minbibnames=1]{biblatex}
\AtEveryBibitem{%
\clearfield{day}%
\clearfield{month}%
\clearfield{endday}%
\clearfield{endmonth}%
}

\newcommand{\ignore}[1]{}
\expandafter\def\expandafter\normalsize\expandafter{%
	\normalsize
	\setlength\abovedisplayskip{6pt}
	\setlength\belowdisplayskip{6pt}
	\setlength\abovedisplayshortskip{6pt}
	\setlength\belowdisplayshortskip{6pt}
}
\DeclareMathOperator*{\argmin}{arg\,min}
\newacronym{psf}{PSF}{Point Spread Function}
\newacronym{snr}{SNR}{Signal-to-Noise Ratio}
\newacronym{bd}{BD}{Blind Deconvolution}
\newacronym{cnn}{CNN}{Convolutional Neural Network}
\newacronym{map}{MAP}{Maximum a Posteriori}
\newacronym{tv}{TV}{Total Variation}
\newacronym{tvrl}{TV-RL}{Total Variation regularized Richardson-Lucy}
\newacronym{rl}{RL}{Richardson-Lucy}
\newacronym{ssim}{SSIM}{Structural Similarity}
\newacronym[plural=AFs,firstplural=Autofocuses (AFs)]{af}{AF}{Autofocus}
\newacronym{sml}{SML}{Sum of Modified Laplacian}
\newacronym{lapv}{LAPV}{Variance of Laplacian}
\newacronym{ewc}{EWC}{Energy of Wavelet Coefficients}
\newacronym{ws}{WS}{Wavelet Sparsity}
\newacronym{gss}{GSS}{Golden Section Search}
\newacronym{sd}{SD}{Standard Deviation}
\newacronym{dof}{DOF}{depth-of-field}
\newacronym{hpf}{HPF}{high-pass filter}
\newacronym{na}{NA}{Numerical Aperture}
\newacronym{roi}{ROI}{Region of Interest}
\newacronym{fov}{FOV}{field-of-view}
\newacronym{fwhm}{FWHM}{full width at half maximum}
\title{{\vspace{-0cm}DeepFocus: a Few-Shot Microscope Slide Auto-Focus using a Sample Invariant CNN-based Sharpness Function}}
\name{Adrian Shajkofci$^{1,2}$, Michael Liebling$^{1,3}$\thanks{This work was funded by the Swiss National Science Foundation, grant 200020\_179217. We thank Cevahir K\"opr\"ul\"u for his work developing the graphical interface of the plugin. The tissue slides were supplied by the EPFL BIOP department from their stock used for training.}}
\address{$^{1}$Idiap Research Institute, CH-1920 Martigny, Switzerland\\
	$^{2}$\'Ecole Polytechnique F\'ed\'erale de Lausanne, CH-1015 Lausanne, Switzerland\\
	$^{3}$Electrical \& Computer Engineering, University of California, Santa Barbara, CA 93106, USA}
\begin{document}
\maketitle
\begin{abstract}
	\gls{af} methods are extensively used in biomicroscopy, for example to acquire timelapses, where the imaged objects tend to drift out of focus. \gls{af} algorithms determine an optimal distance by which to move the sample back into the focal plane. Current hardware-based methods require modifying the microscope and image-based algorithms either rely on many images to converge to the sharpest position or need training data and models specific to each instrument and imaging configuration.
	Here we propose DeepFocus, an \gls{af} method we implemented as a Micro-Manager plugin, and characterize its \gls{cnn}-based sharpness function, which we observed to be  depth co-variant and sample-invariant. Sample invariance allows our \gls{af} algorithm to converge to an optimal axial position within as few as three iterations using a model trained once for use with a wide range of optical microscopes and a single instrument-dependent calibration stack acquisition of a flat (but arbitrary) textured object.
	From experiments carried out both on synthetic and experimental data, we observed an average precision, given 3 measured images, of \SI{0.30 \pm 0.16}{\micro\meter} with a 10$\times$, NA 0.3 objective. We foresee that this performance and low image number will help limit photodamage during acquisitions with light-sensitive samples.

\end{abstract}
\begin{keywords}
Microscopy, autofocus, PSF estimation, convolutional neural networks, Micro-Manager
\end{keywords}

\section{Introduction}
\label{sec:intro}

Modern microscopy techniques rely on many components that are remotely controllable. This allows implementing control loops that limit the need for human-supervised operation. Auto-focusing systems, in particular, are used extensively in the acquisition of timelapses in developmental or cellular biology or to automatically image slides in a slide scanner. In the former application, imaged specimens tend to drift from the focal plane over time because of specimen growth,  flow of the medium, or motion caused by temperature changes. In the latter case, variability in the mounting of the slides requires per-slide adjustment.

\gls{af} systems seek to determine the optimal shift by which to adjust the axial position to maximize image sharpness. \gls{af} solutions can be hardware-based (e.g. laser-based sensing of the sample drift \cite{liron_laser_2006} or phase detection by an auxiliary sensor \cite{silvestri_rapid_2017}) or image-based, which does not require any modification of the optical path of the microscope as a focus score is retrieved from the image itself \cite{sun_autofocusing_2004}.

We can classify image-based \gls{af} algorithms into two categories. The first comprises \gls{af} methods that use iterative minimization of a one-dimensional objective function, the focus score, to move the object to the point at which it is sharpest. Because the output of the function is not predictable and depends on the sample, the \gls{af} has to acquire tens to hundreds of images at different axial positions in order to converge to a non-local optimum \cite{sun_autofocusing_2004}. 
A high number of image acquisitions can be damaging for the sample, especially in fluorescence microscopy \cite{magidson_circumventing_2013}.

Additionally, existing objective functions only give a meaningful result in the neighborhood of the focal plane, and lose information (i.e. the gradient of the curve is zero) farther away from the focal plane. Furthermore, depending on the software implementation and the imaging modality, the acquisition of hundreds of images can take up to several minutes. The second category comprises single shot AF techniques (that need only one or a few images). Thanks to end-to-end \glspl{cnn}, they take an image as input and directly deduce the optimal shift to be in focus (\cite{wei_neural_2018, jiang_transform-_2018, pinkard_deep_2019}). The drawback of these direct methods is that a long and computationally-intensive \gls{cnn} training with a microscope objective-specific training data set, must be repeated whenever the optical system changes. Furthermore, these methods are not directly available in open microscope control software, such as $\mu$Manager \cite{edelstein_computer_2010}.

\begin{figure*}[t!]
	\centering
	\includegraphics[width=0.96\linewidth]{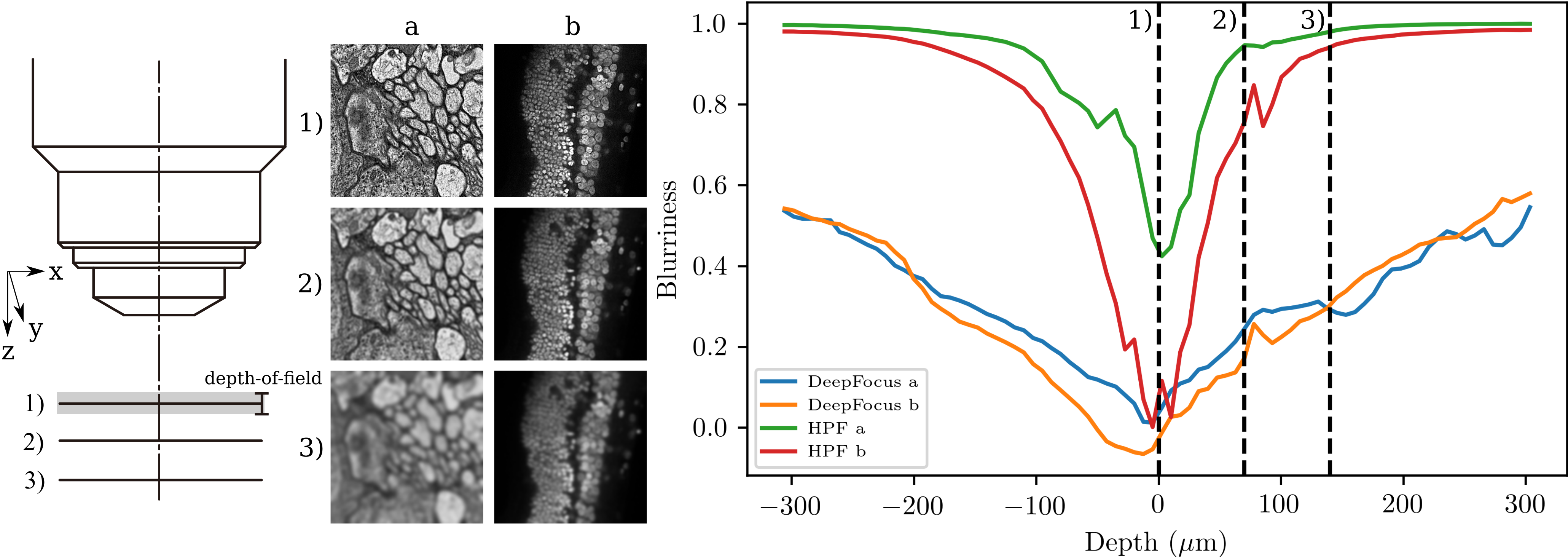}
	\captionsetup{belowskip=-12pt, aboveskip=3pt}
	\caption{The object may be outside of the \acrshort{dof} and appear blurry. Here we quantified the blur $b(z)$ using DeepFocus and an \gls{hpf} for two different images. Using \gls{hpf}, $b(z)$ changes shape and slope when different objects are presented under the microscope, and there is a lack of depth information for $|z| > 150\, \si{\micro\meter}$. Using DeepFocus, the slopes for both images are similar in shape and retain information about depth in the whole $[-300;300]\,\si{\micro\meter}$  region.}
	\label{fig:microscope}
\end{figure*}

In this paper, we propose a local, \gls{cnn}-based focus scoring function that remains nearly invariant when imaging different types of samples or modalities on any given microscope. We developed a correlation-based \gls{af} algorithm that takes advantage of the broad shape and unimodal minimum of this function, which helps to speed up convergence and remaining effective even when the imaged object is far from the focal (several times the \gls{dof}, see Fig. \ref{fig:microscope}). Since our \gls{cnn} method does not require a microscope-specific data set for training besides a single stack of an arbitrary object, it is \emph{plug-and-play}.

This paper is organized as follows. In Section \ref{sec:methods}, we present the blurriness scoring function, the calibration process, and the \gls{af} algorithm. In Section \ref{sec:experiments}, we experimentally verify the scoring function's assumed invariance to a variety of samples and characterize performance with respect to the number of images and in comparison to common \gls{af} scoring functions, using both simulated and experimentally acquired data. We discuss our findings and conclude in Section \ref{sec:discussion}.

\section{Methods}
\label{sec:methods}

\subsection{Problem statement}

We consider a specimen, modeled as 2D manifold in 3D space (such as a thin microscopy slide), that we wish to image with a widefield microscope in bright field, fluorescence, or phase contrast. The entire specimen or some regions in the \gls{fov} can be out of focus and outside of the \gls{dof} (see Fig. \ref{fig:microscope}). We assume the microscope has a motorized stage for adjusting the focus. We aim at finding the optimal axial shift $\Delta z$ by which to adjust the sample position such that it is in focus. We seek a solution that (i) does not require a manually selected reference image to be matched (such that the method can be used both for maintaining focus in live timelapses but also for imaging collections of fixed samples) (ii) requires a minimal number of images (to limit photodamage) (iii) shall not require imaging calibration specimens (PSF measurement beads, etc.) or large-scale, microscope-specific training.

\subsection{Method description}
\label{sec:methods_correlation_based}

The principle behind our proposed algorithm is to measure a blurriness score $b(z_i)$ for a few ($M$) images acquired at different focus positions $z_i$, $i=1,\ldots,M$,  resulting in a set of pairs $\{(z_i,b(z_i)) | i = 1,\ldots, M\}$ and to determine the necessary focal shift $\Delta z$ such that $\{(z_i-\Delta z,b(z_i)) | i = 1,\ldots, M\}$ matches a microscope objective-specific, sample-invariant, depth-blurriness response curve $b_{\text{calib}}(z)$ using cross-correlation. The curve invariance assumption has been similarly used by the model-based curve fitting approach of \cite{yazdanfar_simple_2008}.

For this approach to work, we need a focus estimation function that is invariant to the sample shape or texture (sample-invariance) but co-variant with the sample's axial position and sufficiently informative beyond the immediate vicinity of the focal plane. To this end, we chose an estimator of the local optical properties of the microscope objective \cite{shajkofci_semi-blind_2018}. Briefly, it relies on a trained \gls{cnn} to regress the parameters of a Zernike polynomial \gls{psf} model \cite{von_zernike_beugungstheorie_1934}, given a blurry image patch as an input. Here, we use the estimated Zernike coefficient corresponding to focus as a blurriness score, which provides, given an image as input, a local blurriness score $b[x,y,z]$ for the indicated position depth $z$. 

The trained \gls{cnn} \cite{shajkofci_semi-blind_2018} does not require re-training when used on different microscopes or different microscope objectives and produces a curve whose shape (up to an axial scaling) is invariant to the sample (an aspect that we verify experimentally in Section \ref{sec:experiments_invariance}). In order to determine the axial scaling, which is instrument-dependent, we require a calibration step consisting in the acquisition of a full stack of an arbitrary planar and textured object. This yields a blurriness map $b_{\text{calib}}(z)$ that we center with its minimum at the origin. 

\begin{figure}
		\centering
		\includegraphics[width=0.68\linewidth]{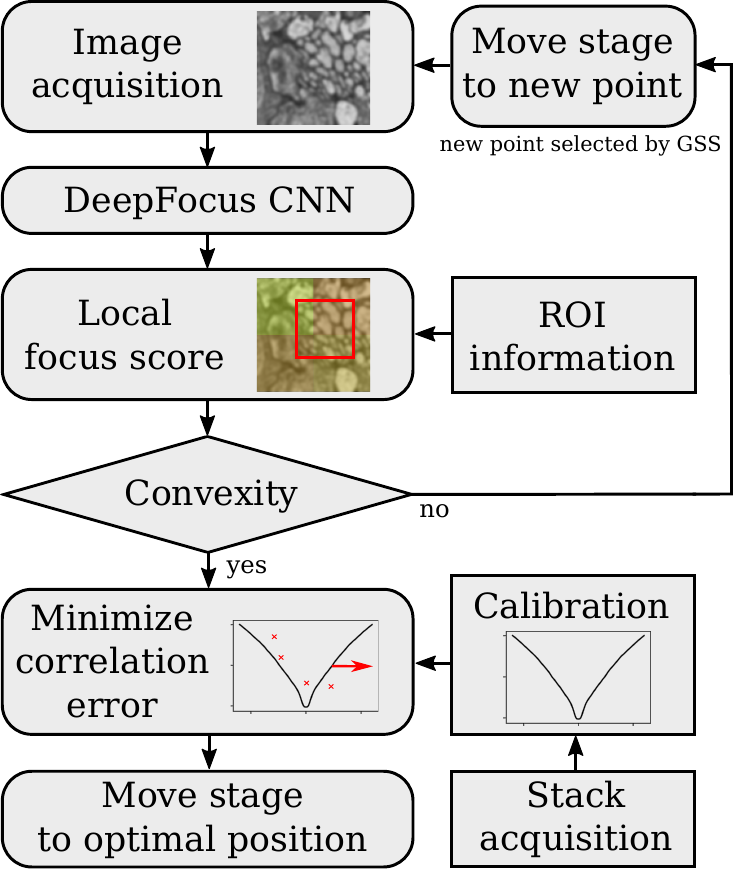}
		\captionsetup{belowskip=-24pt}
		\caption{Flowchart of the AF algorithm (see Section \ref{sec:methods_correlation_based}).}
		\label{fig:flowchart}
\end{figure}

We now describe our proposed \gls{af}, which follows the structure illustrated in Fig.~\ref{fig:flowchart} and is summarized in the steps:
\begin{enumerate}[align=parleft,leftmargin=11pt,labelsep=-8pt,topsep=0pt,itemsep=-1ex,partopsep=1ex,parsep=1ex]
	\item Fit $b_{\text{calib}}$ to a Moffat distribution \cite{moffat_theoretical_1969} and extract its \gls{fwhm}. Set $M = 3$, let $z_1$ be the initial focal plane position, and initialize a \gls{gss} algorithm with the interval $[z_2,z_3]=[z_1 - 2\,\text{\gls{fwhm}},z_1 +2 \,\text{\gls{fwhm}}]$. Acquire images at $z=z_1,z_2,z_3$ and compute, using the \gls{cnn}, the blurriness scores $b[z_1], b[z_2], b[z_3]$.
	\item Check the convexity of $b[z_i]$,  $i = 1, \ldots, M$ by fitting $b[z_i]$ to a quadratic polynomial. If the $R^2$ of the polynomial fit is higher than the $R^2$ of a linear fit, go to Step 6. Otherwise go to Step 3.
	\item Increment $M \mathrel{{+}{=}} 1$. Update the \gls{gss} triplet to obtain and move to a new axial position $z_M$.
	\item Acquire an image at the current axial position $z_M$.
	\item Compute, using the \gls{cnn}, the blurriness score $b[z_M]$ and go to Step 2.
	\item Compute using cross-correlation the local optimal shift $\Delta z(x,y)$ minimizing the squared distance:
	\begin{equation*}
	\Delta z(x,y) = \argmin_{\Delta z} \sum_{i=1}^{M}{  \left(b[x,y,z_i]-b_{\text{calib}}(z_i - \Delta z)\right)^2 }.
	\end{equation*}
	\item Move the sample by ${\Delta z}$, averaged for the \gls{roi} in the $(x,y)$ plane.
\end{enumerate}

\begin{table}[ht!]

		\centering
		\caption{Using the same experimental conditions as Fig. \ref{fig:focusresponse}, we quantified the scoring function performance in terms of \gls{sd} in a  $\SI{100}{\micro\meter}$ range (lower is better) and conditional entropy between the focus score and the distance (lower is better) in the whole $\SI{240}{\micro\meter}$ range. Using experimental acquisitions, DeepFocus outdo all other tested functions in terms of \gls{sd}. Additionally, our method has for both modalities the lower conditional entropy and thus is more informative.}
		\scalebox{0.98}{
			\begin{tabularx}{1.0\linewidth}{ccccc} \toprule
				Focus score function & ${\sigma}_{\text{synth}}$ & ${\text{H}}_{\text{synth}}$ & ${\sigma}_{\text{exp}}$ & ${\text{H}}_{\text{exp}}$\\\midrule
				DeepFocus & $0.03$ & $\bm{1.59}$ & $\bm{0.03}$ & $\bm{5.17}$\\
				\gls{hpf} & $0.06$ & $6.38$  & $0.08$ & $44.31$\\
				Tenengrad \cite{sun_secrets_2010} & $0.10$ & $4.17$  & $0.06$ & $21.77$\\
				\gls{lapv} & $0.03$ & $14.29$  & $0.04$ & $38.28$\\
				\gls{ewc} \cite{hanghang_tong_blur_2004} & $\bm{0.01}$ & $3.47$  & $0.16$ & $8.21$\\
				\gls{sml} & $0.04$ & $7.74$  & $0.06$ & $16.89$\\
				\gls{ws} \cite{liebling_autofocus_2004} & $0.07$ & $5.57$  & $0.19$ & $11.30$\\
				\bottomrule
			\end{tabularx}
		}
		
		\label{fig:information_results}
		\vspace{-12pt}
\end{table}

\section{Experiments}
\label{sec:experiments}

\subsection{Characterization of regression invariance to image diversity}
\label{sec:experiments_invariance}

Since our \gls{af} algorithm relies on the invariance of $b_{\text{calib}}(z)$ to the type of imaged sample, we investigated whether our proposed \gls{cnn} indeed satisfied this condition and whether other (existing) focus metrics could be substituted.

We gathered $N_{\text{synth}}=1000$ images from the evaluation dataset of \cite{shajkofci_semi-blind_2018} and blurred them with Gaussian \glspl{psf} mimicking a 10$\times$, NA $0.3$ objective for $M=132$ points in the depth range $\SI{-120}{\micro\meter} \leq z \leq \SI{120}{\micro\meter}$. In addition, we acquired $N_{\text{exp}}=1000$ stacks of fixed rat brain slices tagged with three fluorescent stains using a widefield transmission light microscope with a 10$\times$, NA 0.3 objective in a depth range of $\SI{-60}{\micro\meter} \leq z \leq \SI{60}{\micro\meter}$. We then computed $b_{\text{calib}}(z)$ using DeepFocus and other methods, including \gls{hpf}, \gls{lapv}, \gls{sml} \cite{nayar_shape_1990}, Tenengrad \cite{sun_secrets_2010}, \gls{ewc} \cite{hanghang_tong_blur_2004}, and \gls{ws} \cite{liebling_autofocus_2004}, which cover a broad range of focus measures, as reviewed in \cite{price_comparison_1994, sun_autofocusing_2004,mateos-perez_comparative_2012,ali_analysis_2018}.

\begin{figure}[ht!]
	\centering
	
	\begin{subfigure}[b]{0.44\linewidth}
		\captionsetup{aboveskip=3pt}
		\centering
		\includegraphics[width=\textwidth]{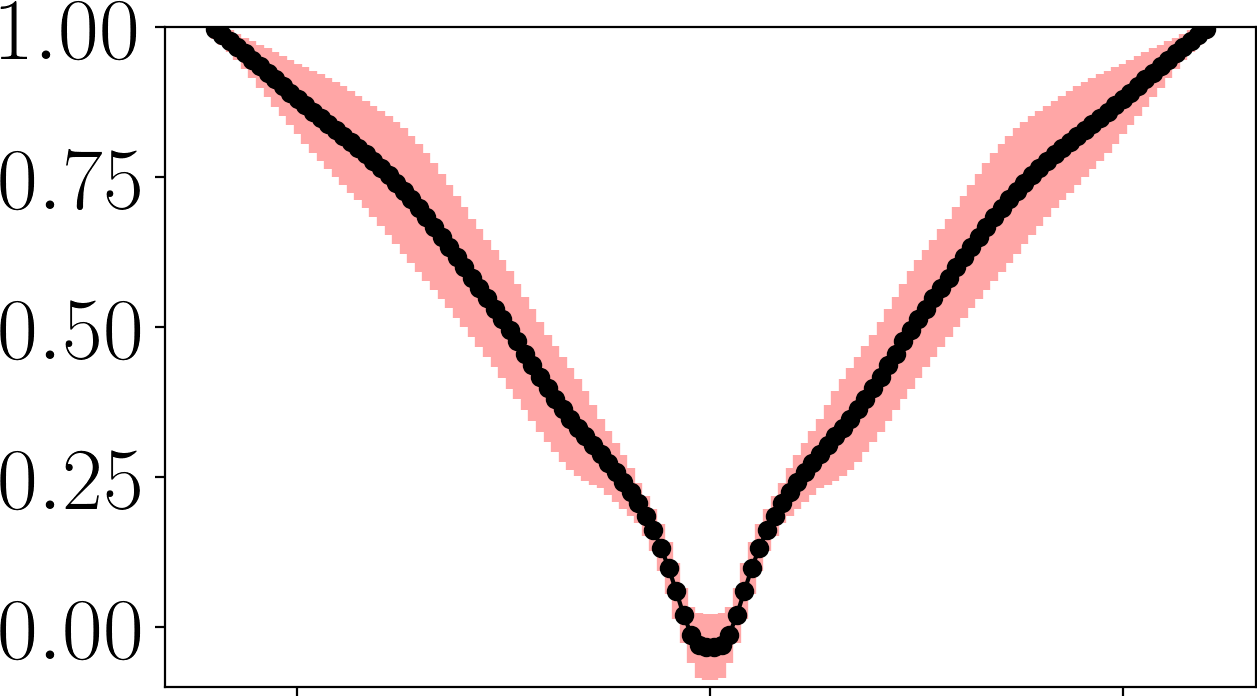}
		\caption{Synt: DeepFocus}
	\end{subfigure}
	\begin{subfigure}[b]{0.44\linewidth}
		\captionsetup{aboveskip=3pt}
		\centering
		\includegraphics[width=\textwidth]{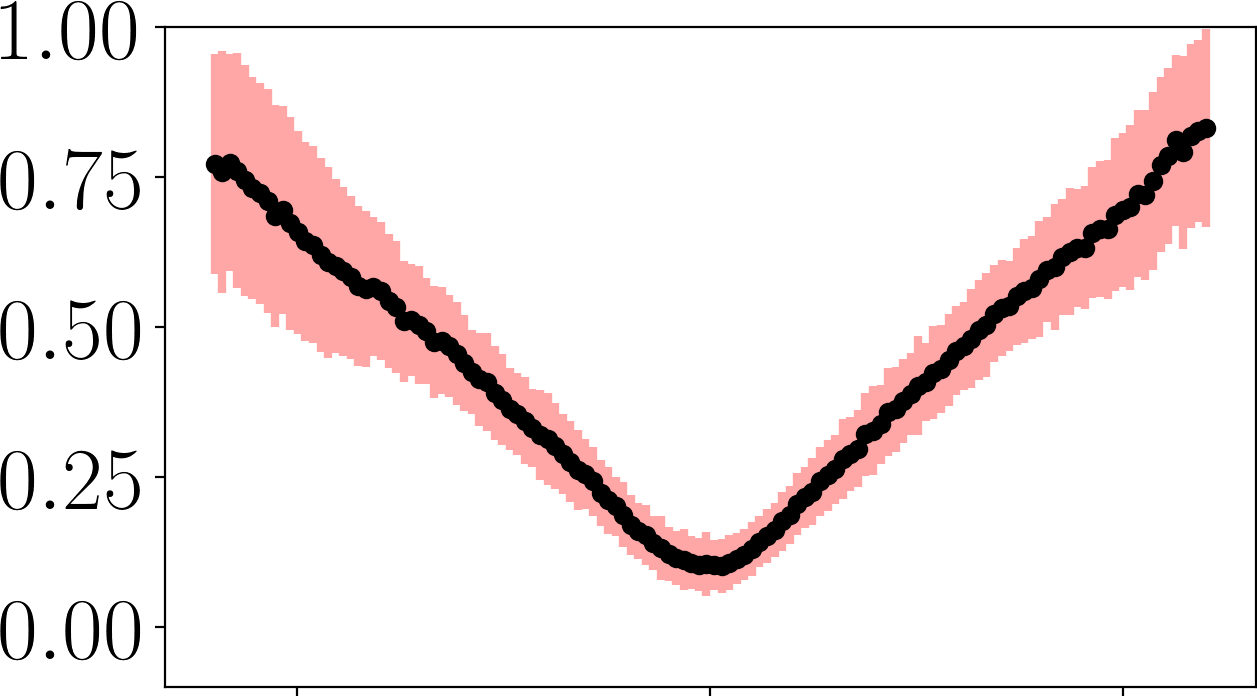}
		\caption{Real: DeepFocus}
	\end{subfigure}
	\hfill
	\begin{subfigure}[b]{0.44\linewidth}
		\captionsetup{aboveskip=3pt}
		\centering
		\includegraphics[width=\textwidth]{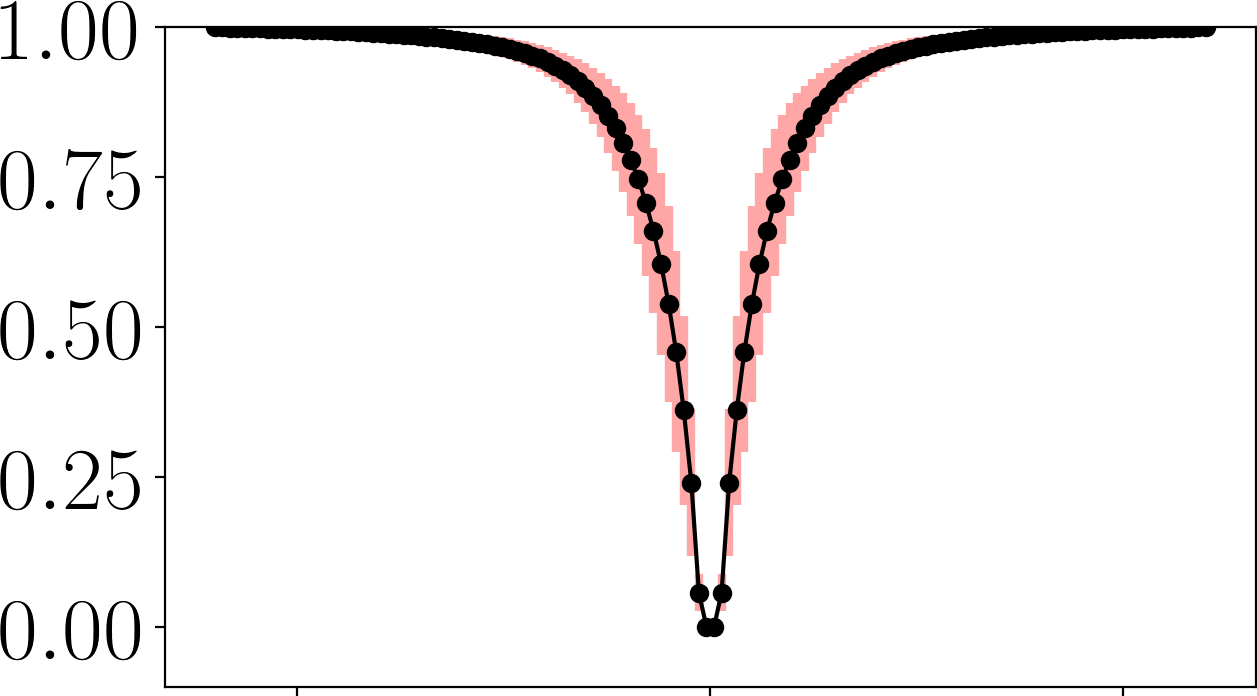}
		\caption{Synt: Tenengrad \cite{sun_secrets_2010}}
	\end{subfigure}
	\begin{subfigure}[b]{0.44\linewidth}
		\captionsetup{aboveskip=3pt}
		\centering
		\includegraphics[width=\textwidth]{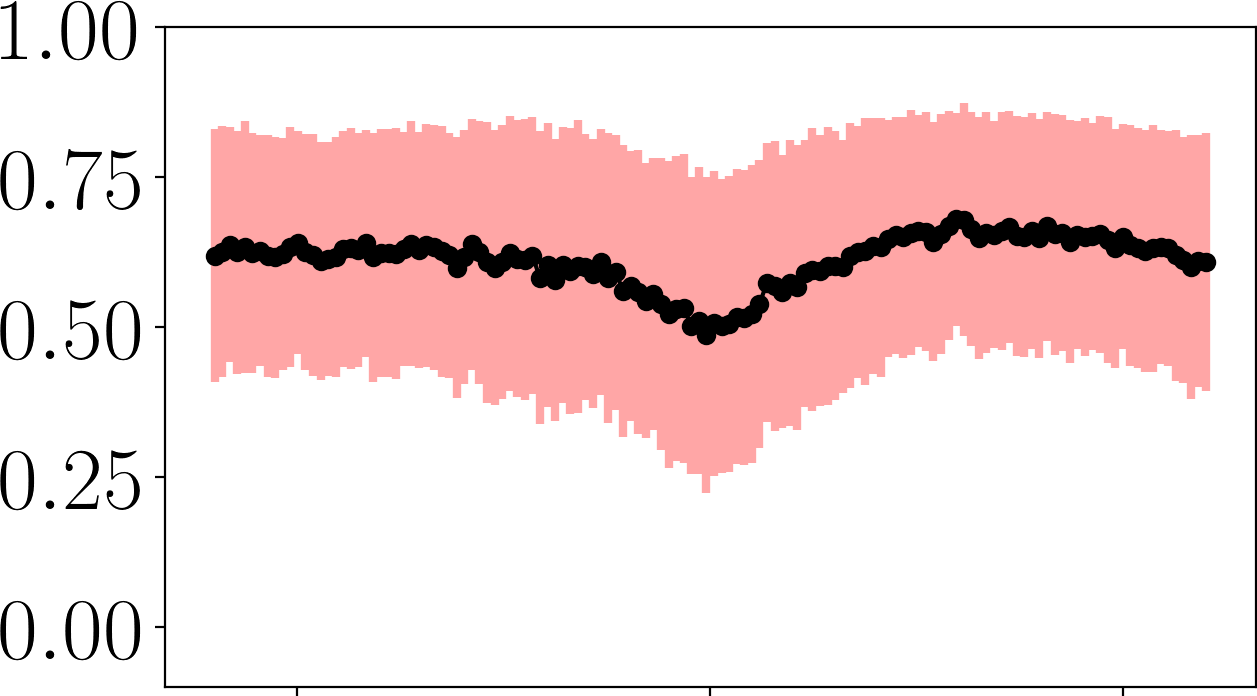}
		\caption{Real: Tenengrad \cite{sun_secrets_2010}}
	\end{subfigure}
	\hfill
	\begin{subfigure}[b]{0.44\linewidth}
	\centering
	\captionsetup{aboveskip=3pt}
	\includegraphics[width=\textwidth]{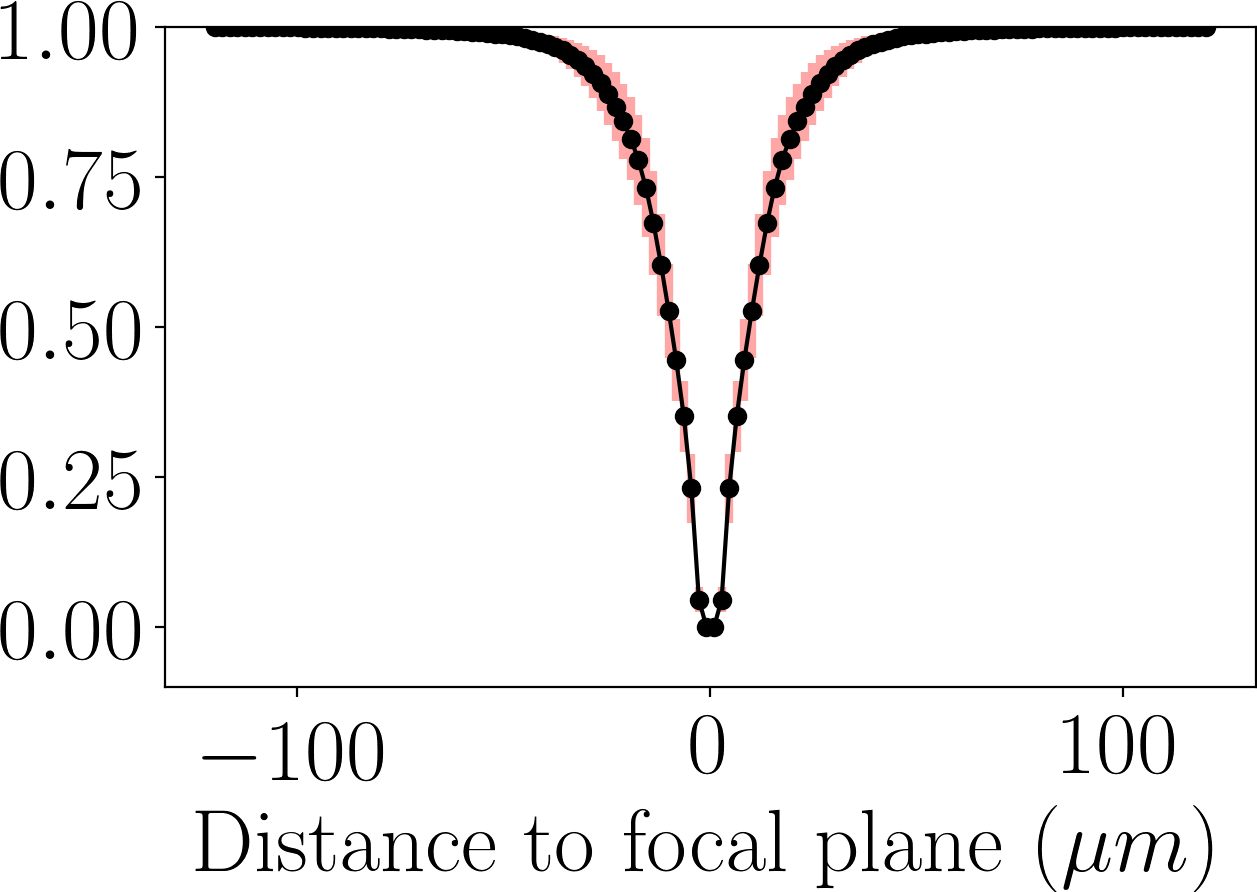}
	\caption{Synt: \gls{ws} \cite{liebling_autofocus_2004}}
\end{subfigure}
\begin{subfigure}[b]{0.44\linewidth}
	\centering
	\captionsetup{aboveskip=3pt}
	\includegraphics[width=\textwidth]{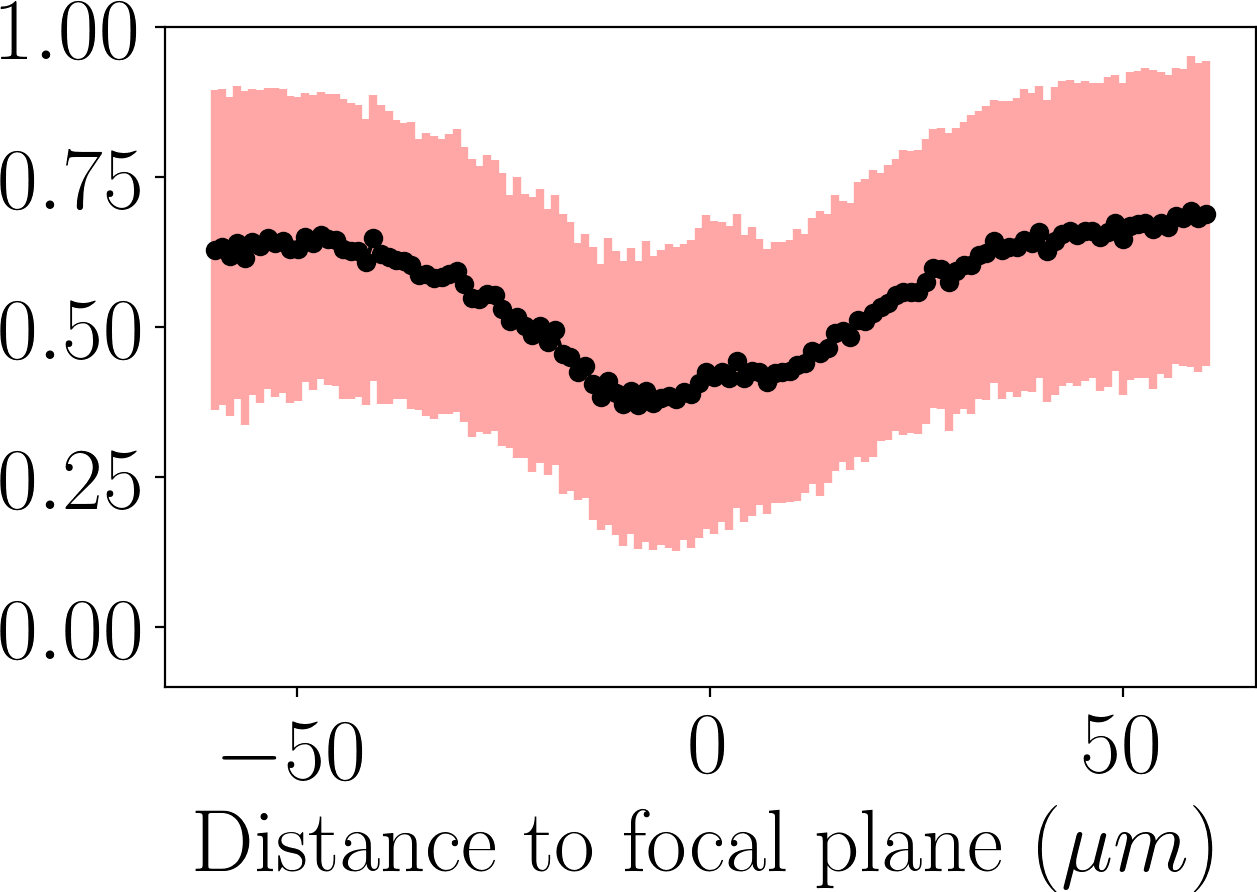}
	\caption{Real: \gls{ws} \cite{liebling_autofocus_2004}}
\end{subfigure}
	\captionsetup{belowskip=-14pt}
	\caption{Comparison of the output of different sharpness scoring functions as a function of $z$, centered at the origin. We used as input synthetically blurred images (left) and stacks of fluorescent rat brain tissue with a 10$\times$, NA 0.3 objective (right). Using DeepFocus, the \gls{sd} of  $b_{\text{calib}}(z)$ around the focal plane is lower than with the other scoring functions. Additionally, scoring functions other than DeepFocus do not infer depth information ($b_{\text{calib}}(z)=1$ for all $z$) when $|z| > 50 \si{\micro\meter}$.}
	\label{fig:focusresponse}
\end{figure}

In Table \ref{fig:information_results}, we reported the average \gls{sd} of $b_{\text{calib}}(z)$ over all input images. Using the experimental dataset, our method had an average \gls{sd} of $\sigma = 0.03$ (normalization scale with 1 and 0 the blurriest and sharpest values, respectively). We noticed, as illustrated in Fig. \ref{fig:focusresponse} (a) and (b), that DeepFocus' \gls{sd} increased when $|z|$ increases (i.e when the acquired pictures contain a medium-to-high blur). A low \gls{sd} implies that $b_{\text{calib}}(z)$ is similar with different types of imaged specimens. Other methods had a \gls{sd} of $0.04 < \sigma <0.19$, and hence confirmed the variance of these focus metrics with image diversity.

\subsection{Characterization of information measure of the scoring function}
\label{sec:experiments_invariance_information}

We next investigated how robustly our proposed DeepFocus measure can report (de)focus information as the distance from focus is increased up to 10 times the \gls{dof}. 
We observed (Fig.~\ref{fig:focusresponse}) that focus metrics other than ours were unable to give any information about $z$ from $b_{\text{calib}}(z)$ whenever $|z|$ is higher than \SI{60}{\micro\meter}, as they reach a value that does no longer vary as the position is increased further. Since the gradient in such plateau regions is small, minimization algorithms could not converge quickly. To quantify these visual observations regarding the uncertainty of recovering $z$ from any given $b_{\text{calib}}(z)$, we computed the conditional entropy:
\begin{equation*}
H(\mathbf B | \mathbf Z) = - \hspace{-1em}\sum_{b_{\text{calib}}  \in \mathcal B, z \in \mathcal Z}\hspace{-1em}{p(b_{\text{calib}} | z) \log\left(\frac{p(b_{\text{calib}}  | z)}{\sum_{z_i}{p(b_{\text{calib}}  | z_i)}}\right)},
\end{equation*}
where $\mathbf B$ and $\mathbf Z$ are random variables representing the calibration blurriness score and the axial distances, $\mathcal B$ and $\mathcal Z$ their support sets, and $p(b_{\text{calib}}  | z)$ the probability of a score $b_{\text{calib}} $, given the distance $z$. A high conditional entropy value implies a high uncertainty of detecting the right $z$ position for a given $b_{\text{calib}} $. The results, compiled in Table \ref{fig:information_results}, reveal that DeepFocus had a conditional entropy of $H_{\text{synth}}= 1.59$, a value smaller than that obtained when using any of the other scoring functions instead. In the case of experimental acquisitions, we observed again an improvement in terms of entropy ($H_{\text{exp}}= 5.17$), where other methods have values in the range $8.21 < H_{\text{exp}} < 44.31$.
We further determined the threshold distance after which no distance information can be inferred from the image, i.e when the image is too blurry to make the \gls{af} converge. DeepFocus retained depth information for a range of $120$ $\mu$m with a 10$\times$, $\text{NA}=0.3$ objective, which is equivalent, using the diffraction-limited \gls{dof} formula, to 11 times the \gls{dof} (\SI{10.7}{\micro\meter}). In comparison, metrics like \gls{ws} and \gls{sml} achieved ranges of only 4 and 7 times the \gls{dof}, respectively.

\subsection{Characterization of the \gls{af} error as a function of the number of acquisitions}

\begin{figure}

		\centering
		\includegraphics[width=0.95\linewidth]{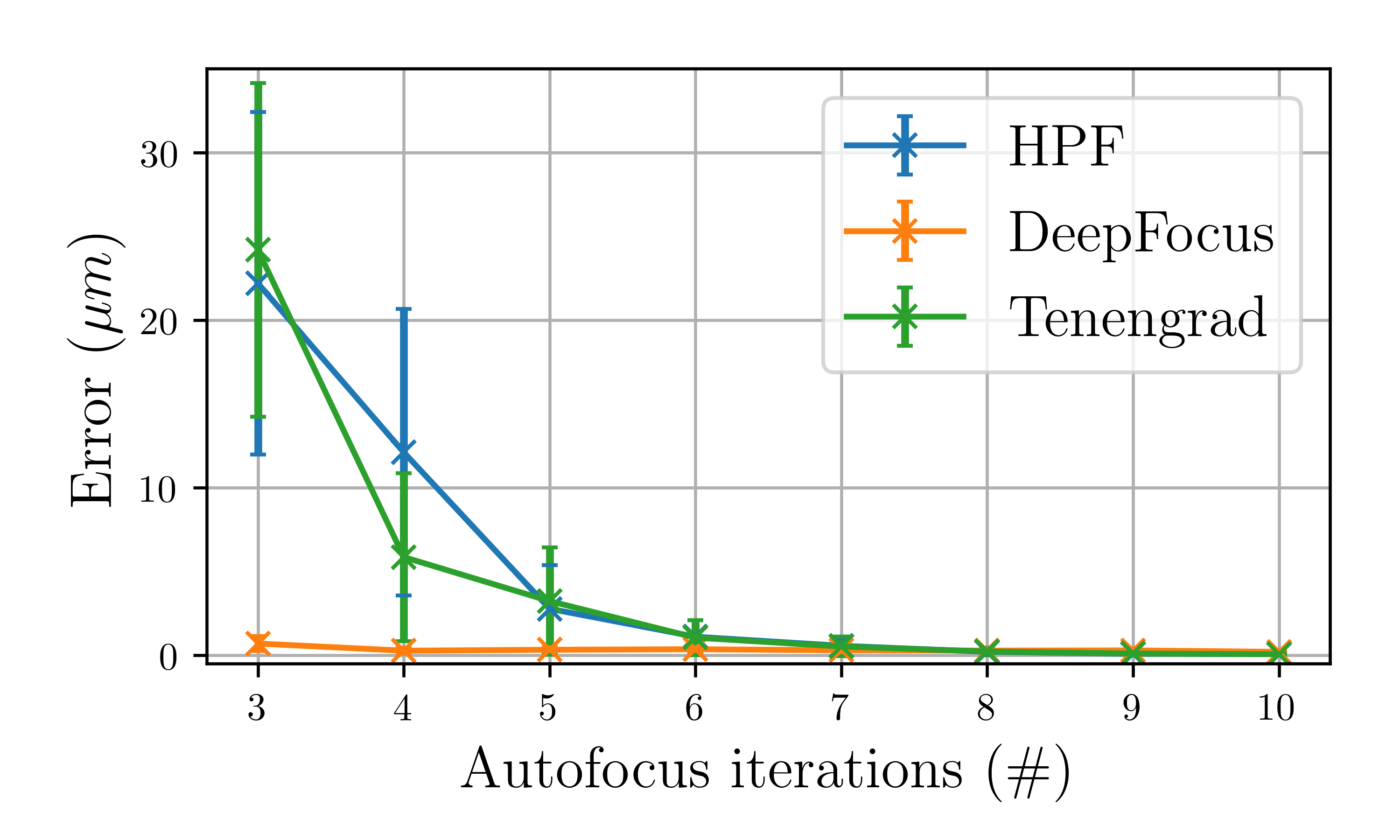}
		\captionsetup{belowskip=-18pt, aboveskip=0pt}
		\caption{Comparison of the \gls{af} error using 3 different \gls{af} scoring functions for 100 samples. We quantified the distance between the theoretical focus plane position and the \gls{af} output as a function of the number of \gls{af} iterations which  represent additional input images. DeepFocus yields an error of \SI{0.27 \pm 0.18}{\micro\meter} with 4 iterations. With 8 iterations or more, the others methods are on par or more accurate than ours.}
		\label{fig:number_images}

\end{figure}

We finally investigated how accurately DeepFocus could retrieve the focal distance as a function of the number of images acquired. We used 100 blurred images from the generated dataset in Section \ref{sec:experiments_invariance} with a known in-focus position and computed its distance to the output position of the \gls{af}. We also compared our method to other autofocus scoring functions (for which we used a bounded Brent's method as optimizer). The results are summarized in Fig.~\ref{fig:number_images}.

We observed that our proposed \gls{af} converged rapidly (3 iterations), while the two other focus functions needed more than twice as many images to reach a similar focus accuracy.  
Using 8 iterations or more, we did not notice a better accuracy with our method compared to Tenengrad or \gls{hpf}.

\section{Discussion and conclusion}
\label{sec:discussion}

In our experiments, we showed that the variance of $b_{\text{calib}}$ over multiple images was usually lower using DeepFocus than when using other focus scoring functions, especially near the focal plane. Our explanation would be that the \glspl{cnn}, already known to be translation-invariant \cite{lecun_learning_2012}, have been trained specifically for the recognition of the \gls{psf} parameters without discrimination on the input image type and position. By contrast, crafted features such as \glspl{hpf} are computed from content-based calculations and differ from one image to another.
When the image is acquired at a large distance from the focal plane, we noticed a loss of spatial features in the acquired image, due to the large \gls{fwhm} of the \gls{psf} that degraded it. However, we have been able to retrieve depth information from the image up to 2.5 times farther away from the focal plane than with other methods. That could be mostly explained by the fact that DeepFocus computes features from a 128$\times$128 px window, while Gradient-based methods use a much smaller window, such as 3$\times$3 or 5$\times$5.

In summary, we developed an \gls{af} method based on a combination of an \gls{cnn} scoring function and optimization algorithms that are relying on the invariance of the scoring function. We showed that DeepFocus was robust to changes amongst samples, which enables the retrieval of the optimal axial shift using a correlation-based optimization process that needs as few as 3 images to converge. 
Our method is currently limited to imaging thin samples and further work will investigate the procedure for thicker objects. We implemented the calibration step and \gls{af} algorithm as two plugins (Java with a PyTorch \cite{paszke_automatic_2017} backend) for the $\mu$Manager microscopy acquisition engine \cite{edelstein_computer_2010}, which we will make available upon acceptance.

\section{References}

\small{
\begin{spacing}{1.0}

\printbibliography[heading=none]

\end{spacing}
}
\end{document}